\documentclass[aps,pra,reprint,onecolumn,notitlepage,eqsecnum,floatfix,showkeys,nofootinbib]{revtex4-2}

\usepackage{amsmath}
\usepackage{xcolor}
\usepackage{graphicx}
\usepackage{hyperref}
\usepackage{multirow}
\usepackage{linearb}
\usepackage{relsize}

\newcommand{\bq}{\begin{eqnarray}}
\newcommand{\eq}{\end{eqnarray}}
\newcommand{\bqn}{\begin{eqnarray*}}
\newcommand{\eqn}{\end{eqnarray*}}
\newcommand{\bqs}{\begin{subequations}}
\newcommand{\eqs}{\end{subequations}}
\newcommand{\bw}{\begin{widetext}}
\newcommand{\ew}{\end{widetext}}

\newcommand{\aaa}{{\boldsymbol a}}
\newcommand{\bbb}{{\boldsymbol b}}

\newcommand{\qq}{{\boldsymbol q}}

\newcommand{\LL}{{\boldsymbol L}}
\newcommand{\pp}{{\boldsymbol p}}

\DeclareRobustCommand{\looongrightarrow}{%
  \DOTSB\relbar\joinrel\relbar\joinrel\relbar\joinrel\rightarrow
}

\newcommand{\red}[1]{{#1}}
\newcommand{\green}[1]{{#1}}
\newcommand{\blue}[1]{{#1}}
\newcommand{\orange}[1]{{#1}}
\newcommand{\brown}[1]{{#1}}

\begin{document}
\title{Coherent State Path Integral Monte Carlo}

\author{Riccardo Fantoni}
\email{riccardo.fantoni@scuola.istruzione.it}
\affiliation{Universit\`a di Trieste, Dipartimento di Fisica, strada
  Costiera 11, 34151 Grignano (Trieste), Italy}

\date{\today}

\begin{abstract}
\blue{We propose a new quantum simulation method for a many body quantum liquid of identical particles
at finite (non-zero) 
temperature. The new scheme expands the high temperature density matrix on the overcomplete
set of single particles coherent states of John Rider Klauder instead of the usual plane waves 
as in conventional path integral methods. One is free to tune the elastic constant 
and/or the mass of the harmonic oscillator subtending the coherent states so as to maximize 
the computational efficiency of the algorithm. We prove that in the limit of an extremely stiff 
harmonic oscillator the results for the internal energy tends towards the correct expected values. 
Moreover we suggest that a stiff harmonic oscillator could allow the use of larger (imaginary)
timesteps. This additional degree of freedom is the characteristic feature of our new algorithm 
and is not available in more conventional path integral methods.}
\end{abstract}


\maketitle
\section{Introduction}
\label{sec:intro}

\orange{The Path Integral Monte Carlo method has been used in statistical physics by a long time 
\cite{Fosdick1966}. For fermionic systems it was worked out by V. S. Filinov 
\cite{Filinov1975}. A recent review by T. Dornheim et al. \cite{Dornheim2018} summarizes 
the use of path integral Monte Carlo method for fermionic systems. Whereas Boltzmann
and Bose particles can be solved exactly at least computationally, Fermi particles cannot 
due to the notorious {\sl sign problem} of Feynman 
\cite{Ceperley1991,Ceperley1996,Feynman1982,Feynman-Hibbs}. The sign problem for the 
ideal Fermi-gas can be solved using the so-called ``Brownian bridge'' \cite{Filinov2020}.}

In this work we describe a new algorithm able to simulate a quantum liquid made of identical particles
at finite temperature through the cooperation of coherent states (CS) 
\cite{KlauderJCP1963a,KlauderJCP1963b,KlauderJCP1963c,KlauderJCP1963d,KlauderJCP1963e,Klauder-CS} and the 
Path Integral Monte Carlo (PIMC) method \cite{Ceperley1995}. The algorithm, that we will call Coherent 
States Path Integral Monte Carlo (CSPIMC), reconstructs the equilibrium hot thermal density matrix of a 
many body system of particles at each small imaginary time step thanks to the properties of the single 
particle coherent states that form an overcomplete set 
\cite{KlauderJCP1963a,KlauderJCP1963b,KlauderJCP1963c,KlauderJCP1963d,KlauderJCP1963e,Klauder-CS}. The 
coherent state is a state of minimal uncertainty which is defined to be the (unique) eigenstate of the 
annihilation operator of a Harmonic Oscillator (HO) and as such it is described by a wave function 
whose probability distribution is a Gaussian. The information on the thermal density matrix 
after a sufficiently big number of sufficiently small imaginary time steps $\tau$, so to reach the 
desired finite inverse temperature $\beta$, is then reconstructed into a path integral through the PIMC 
calculation. As usual we take $\beta=1/k_BT=M\tau$ with $k_B$ Boltzmann constant, $T$ the absolute 
temperature, $M$ the number of time steps discretizations between 0 and $\beta$, \blue{and $\tau$ a 
time discretization ultraviolet cutoff.}

We suggest that this way of simulating a quantum many body system of bosons/fermions may allow
to choose larger timesteps $\tau$ which will make the Monte Carlo algorithm more efficient. This
will not solve the \red{sign problem of Feynman 
\cite{Ceperley1991,Ceperley1996,Feynman1982,Feynman-Hibbs} which is 
still an open problem in statistical physics}. In particular we see, as is clearly shown in Appendix 
\ref{app:zeta}, that choosing the HO mass big enough allows to work at larger timesteps.

Our novel Quantum Monte Carlo (QMC) algorithm adds to the rich variety of similar methods for a finite 
temperature numerical experiment starting from the conventional {\sl standard simulations} of D. M. 
Ceperley \cite{Ceperley1995}, passing to the {\sl worm-algorithm} of M. Boninsegni \cite{Boninsegni2006a}, 
to end to the {\sl pair-product approximation} used by E. W. Brown \cite{Brown2013}. All these Monte Carlo 
methods are based on the Metropolis algorithm \cite{Metropolis,Kalos-Whitlock} \green{and use plane waves 
to expand the kinetic energy action}.

If, from one side, we do not need to specify further the {\sl primitive approximation} 
\cite{Ceperley1995} for the {\sl potential energy action}, \green{diagonal in position 
representation}, on the other side, we will make a rather brute force 
approximation for the {\sl kinetic action} that hinges on the peculiar properties of coherent 
states. 
\red{We then propose a ``thought'' computer experiment with no ambition to carry out
our own numerical experiment for fermions. But we successfully tested the new proposed kinetic 
energy {density matrix} on simulations of distinguishable Boltzmann \green{and identical Bose free} 
particles.}
\orange{These computer experiments show the approach to the correct thermodynamic value for 
the \brown{thermodynamic} estimator of the kinetic energy in
the limit of a large value for $\xi=m_{h.o.}\omega/2$ where $m_{h.o.}$ is the HO mass and 
$\omega$ is its frequency. For example we performed a test simulation of 16 $^4He$ atoms near the 
continuum limit ($M=250$) and compared our CSPIMC result with the one of conventional 
plane waves PIMC for both the kinetic energy and the potential energy. 
The freedom of choice of the stiffness of the HO is the characteristic 
feature of our new algorithm that is not present in more conventional path integral 
approaches. This freedom of choice of the stiffness of the HO, regulated by
its mass, $m_{h.o.}$, and elastic constant, $k=m_{h.o.}\omega^2$ goes together with the 
freedom of choice of the mass, $m$, of the non interacting particles. For the 
interacting system the mass $m$ plays a role and one can renormalize the strength of the 
interaction, the coupling constant, $\upsilon\to m\bar{\upsilon}$, and work at fixed 
$\bar{\upsilon}$. We will see how the two masses are mingled together in the adimensional 
constant $\varphi=\xi\tau/m$ that naturally appears in our formulation.}

\section{The algorithm}
\label{sec:algo}

Let us consider a many body system of $N$ particles with positions 
$Q=(\qq_1,\qq_2,\ldots,\qq_N)$ and momenta $P=(\pp_1,\pp_2,\ldots,\pp_N)$ in 
thermal equilibrium at a finite temperature $T$ in a volume $\Omega$ at a density
$\rho=N/\Omega$.

The equilibrium statistical mechanics description of the many body fermions requires 
the knowledge of the thermal density matrix operator $\hat{\rho}=\exp(-\beta\hat{H})$ 
where $\hat{H}$ is the fermions Hamiltonian operator, $\beta=1/k_BT$ is the ``inverse 
temperature'', and $k_B$ is the Boltzmann constant.

The thermal density matrix satisfies to the Bloch equation
\bq
\frac{\partial\hat{\rho}}{\partial\beta}=-\hat{H}\hat{\rho}.
\eq

If we know the eigenstates and eigenvalues of the Hamiltonian, $|\Psi_i\rangle$ 
and $E_i$, we can use the completeness of this system of orthonormal states  
to write the position representation of the density matrix as follows
\bq \label{eq:pdm}
\rho(Q,Q';\beta)=\langle Q|\hat{\rho}|Q'\rangle=\sum_i\langle Q|\Psi_i\rangle 
e^{-\beta E_i}\langle\Psi_i| Q'\rangle.
\eq
Otherwise, in the high temperature limit we can neglect terms of orders higher than one in the
small $\tau$ in the Baker-Campbell-Hausdorff formula
\footnote{\orange{This is called the {\sl primitive approximation} and is valid whenever the 
kinetic energy and the potential energy operators are separately bounded from below which is 
not the case for a two-component Coulomb liquid for which it is necessary to use the pair 
product density matrix as the building block \cite{Pollock1988}.}} 
to find \cite{Ceperley1995}
\bq \label{eq:pa}
\rho(Q,Q';\tau)=\langle Q|e^{-\tau\hat{H}}|Q'\rangle\approx
\langle Q|e^{-\tau\hat{T}}e^{-\tau\blue{\upsilon}\hat{V}}|Q'\rangle,
\eq
where $\hat{H}=\hat{T}+\blue{\upsilon}\hat{V}=\hat{P}^2/2m+\blue{\upsilon}V(Q)$ is the Hamiltonian 
operator with $\hat{T}$ the kinetic energy operator, $\hat{V}$ the potential energy operator, 
\blue{with $\upsilon$ the coupling constant}, $\hat{Q}=Q$ the positions 
operator, $\hat{P}=-i(\nabla_{\qq_1},\nabla_{\qq_2},\ldots,\nabla_{\qq_N})$ the momenta operator, 
and $m$ the particles mass; here and in the following we choose $\hbar=1$.

Taking $\tau=\beta/M$ with $M$ a large integer we can then reconstruct the finite 
temperature density matrix using Trotter formula in the following successive `convolutions' 
\cite{Trotter1959,Ceperley1995} as usual
\bq \label{eq:trotter}
\rho(Q,Q';\beta)=\int\rho(Q,Q_1;\tau)\cdots\rho(Q_{M-1},Q';\tau)\,dQ_1\cdots dQ_{M-1}.
\eq

Since 
$\hat{\rho}\approx e^{-\tau\hat{T}}e^{-\tau\blue{\upsilon}\hat{V}}=\prod_\alpha e^{-\tau\hat{T}_\alpha}e^{-\tau\blue{\upsilon}V}$, 
where $\hat{T}_\alpha$ is the kinetic energy of particle $\alpha$ and the exponential 
containing the potential is diagonal in position space and just a multiplicative factor, 
then the many body state $|\Upsilon_a\rangle$ factorizes into a product 
of single particle states $\prod_\alpha|\psi_\alpha^a\rangle$ 
\bq
|\Upsilon_a\rangle=\prod_{\alpha=1}^N|\psi_\alpha^a\rangle,
\eq
where $a$ labels the set of many body states which inherit the overcompleteness of the 
single particles states. Here and in the following we will use Greek indexes to denote
the particle label and Latin indexes to denote the label of the (over)complete single 
particle states.

Antisymmetrizing so to satisfy Fermi statistics \orange{(or symmetrizing so to satisfy Bose 
statistics)}, we find
\bq \label{eq:spe}
|\Upsilon_a\rangle\langle\Upsilon_a|=\frac{1}{N!}\sum_{\cal P}(-1)^{\cal P}\prod_{\alpha,\beta=1}^N
|\psi_\alpha^a\rangle\langle\psi_{{\cal P}\beta}^a|
=\frac{1}{N!}{\rm det}|| \,|\psi_\alpha^a\rangle\langle\psi_{\beta}^a|\, ||,
\eq
where ${\cal P}$ is any of the $N!$ permutations of the $N$ particles.

Now we can take as the single particle states $|\psi_\alpha^a\rangle$ the coherent states 
\cite{KlauderJCP1963a,KlauderJCP1963b,KlauderJCP1963c,KlauderJCP1963d,KlauderJCP1963e,Klauder-CS}
\bq
|\psi_\alpha^a\rangle\equiv e^{-i\qq_a\cdot\hat{\pp}_\alpha}e^{i\pp_a\cdot\hat{\qq}_\alpha}|0\rangle, 
\eq
where $|0\rangle$ is the ground state of the three dimensional harmonic oscillator of elastic 
constant $k$ along all three dimensions
\footnote{\orange{
Our discussion will be carried out in three dimensions for definiteness but the generalization to any 
dimensions is straightforward.}}.
The coordinate representation of this state is
\bq \label{eq:csq}
\psi^\aaa(\qq_\alpha)\equiv\langle\qq_\alpha|\qq_a,\pp_a\rangle&=&\left(\frac{m_{h.o.}\omega}{\pi}\right)^{3/4}e^{-\frac{m_{h.o.}\omega}{2}\left[\qq_\alpha-\sqrt{\frac{2}{m_{h.o.}\omega}}{\rm Re}(\aaa)\right]^2+i\qq_\alpha\cdot\sqrt{2m_{h.o.}\omega}{\rm Im}(\aaa)-i2{\rm Re}(\aaa)\cdot{\rm Im}(\aaa)},\\
\aaa&=&\frac{1}{\sqrt{2m_{h.o.}\omega}}(m_{h.o.}\omega\qq_a+i\pp_a),
\eq
where $\omega=\sqrt{k/m_{h.o.}}$ is the angular frequency of the harmonic oscillator of elastic 
constant $k$ and mass $m_{h.o.}$. \blue{Note that the HO, or {\sl ghost}, coordinates, $\qq_a$, 
are elastically bound to the particle, or {\sl real}, coordinates, $\qq_\alpha$, by a gaussian
and viceversa. 
In other words the HO ``$a$'' is centered on the particle ``$\alpha$''. The width of 
the gaussian is proportional to $(m_{h.o.}\omega)^{-1}$ so that in the $m_{h.o.}\omega\to\infty$ 
limit the ghost coordinates coincide with the particle coordinates.}

This way we obtain the thermal density matrix at a finite inverse temperature 
$\beta$ through the ``multiple convolution'' integral (\ref{eq:trotter}), but with
the following high temperature density matrix
\footnote{\orange{A general expression for the high-temperature density matrix was obtained by 
G. Kelbg in 1963 \cite{Kelbg1963}. This expression contains the Fourier transform of the
interaction potential, so it is possible to obtain high-temperature density matrices for a 
variety of potentials (for example, it is possible to take into account long-range interactions).}}
\bq \label{eq:K}
\rho(Q,Q';\tau)&\approx &e^{-K(Q,Q';\tau,m,\xi)}e^{-\tau\blue{\upsilon}V(Q')}\\ \label{eq:K1}
&=&e^{-\tau\blue{\upsilon}V(Q')}\frac{1}{N!}\sum_{\cal P}(-1)^{\cal P}\prod_{\alpha}\zeta_\alpha\Bigl[\qq_{{\cal P}\alpha}|\qq'_\alpha;\tau,m,\xi\Bigr],
\eq
where $K$ is the kinetic part of the semiclassical action depending on the expansion of 
$|\Upsilon_a\rangle$ on the single particle coherent states of Eq. (\ref{eq:spe}), 
$\zeta$ is defined in Eq. (\ref{eq:zeta}) in Appendix \ref{app:K} and its primitive 
approximation is determined in Eqs. (\ref{eq:azeta}), of Appendix \ref{app:zeta}, 
$k$ is the elastic constant of the harmonic oscillator, and $m_{h.o.}$ is its mass.

So that in the $M\to\infty$ limit the Trotter formula (\ref{eq:trotter}) becomes a path 
integral made of the $M$ high temperature density matrices (\ref{eq:K1}) at each time step
\cite{Trotter1959}.

Note that if we choose an extremely stiff harmonic oscillator, i.e. one such that
 $m_{h.o.}\omega\to\infty$, then the Gaussian 
$|\psi^a_\alpha(\qq)|^2$ of Eq. (\ref{eq:csq}) reduces to a Dirac $\delta$ centered on the 
position $\qq$ only. We suggest that this may allow the use of larger timesteps in Eq. 
(\ref{eq:trotter}) that reduces to the path integral. And this will increase the efficiency of 
the MC algorithm.

As usual in order to measure an observable $\hat{{\cal O}}$ we need to calculate 
$\langle\hat{{\cal O}}\rangle={\rm tr}(\hat{\rho}\hat{{\cal O}})/{\rm tr}(\hat{\rho})$.
This requires to impose periodic boundary conditions on the imaginary time so that 
$\rho(Q,Q';t)=\rho(Q,Q';t+\beta)$.

Moreover in a simulation we want to mimic the thermodynamic limit as close as possible and 
this is usually obtained enforcing spatial periodic boundary conditions juxtaposing an 
infinite number of identical copies of the simulation box of volume $\Omega=L_1L_2L_3$ along 
the three dimensions. This can be easily obtained by taking for each particle 
$\qq_\alpha+\LL=\qq_\alpha$, i.e a periodic box. Of course as $\Omega$ increases we will mimic the 
thermodynamic limit closer and closer. One usually refers to this feature of a computer experiment 
as the {\sl finite size error}. This can be obtained with the expansion in coherent states by taking
the following infinite sum
\footnote{Given any function $f(x)$ it can always be made periodic of period $L$ by choosing 
$f_L(x)=\sum_{n=-\infty}^\infty f(x+nL)$.} at the end
\bq \label{eq:zp}
\zeta_\alpha\to\zeta_\alpha^L=\sum_{i,j=-\infty}^\infty\zeta_\alpha\Bigl[\qq+iL|\qq'+jL;\tau,m,\xi\Bigr],
\eq
where we assumed $L_1=L_2=L_3=L$ for simplicity.
\footnote{\orange{Note that the series (\ref{eq:zp}) may be conditionally convergent.}}

\red{
\section{Comparison with the standard kinetic energy {density matrix}}
\label{sec:comparison}

We successfully performed some preliminary test simulations with our new high temperature
{\sl coherent states} {density matrix} of Eq. (\ref{eq:K}) and compared it with the standard 
{\sl plane waves} primitive approximation \citep{Ceperley1995} with
\bq \label{eq:Kpw}
K(Q,Q';\tau,m)=\frac{(Q-Q')^2}{4\lambda\tau}+\frac{3N}{2}\ln(4\pi\lambda\tau),
\eq
and $\lambda=1/2m$.

In particular our proposed new path integral expression can be written as 
\bq \label{eq:eureka}
&&\rho(Q_0,Q_\beta;\beta)=\\ \nonumber
&&\frac{1}{N!}\sum_{\cal P}(-1)^{\cal P}\mathlarger{\int}_{\mathsmaller{\substack{Q_0\to Q(t)\to{\cal P}Q_\beta}}} e^{-\mathlarger{\int}_0^\beta \frac{P_a(t)^2}{2m}+\blue{\upsilon}V[Q(t)]\,dt} \substack{\beta\\\mathlarger{\mathlarger{\mathlarger{\mathlarger{\textlinb{\Bo}}}}}\\t,t'=0}\prod_{\alpha=1}^N \psi_\alpha^a[\qq(t)]{\psi_\alpha^b}^*[\qq(t')]G_{a,b}\;\frac{{\cal D}Q_a{\cal D}P_a}{(2\pi)^{3N}}\;\frac{{\cal D}Q_b{\cal D}P_b}{(2\pi)^{3N}}\;{\cal D}Q(t),
\eq
where $Q_a=({\qq_a}_1,\ldots,{\qq_a}_N)$ and 
$P_a=({\pp_a}_1,\ldots,{\pp_a}_N)$, $Q_b=({\qq_b}_1,\ldots,{\qq_b}_1)$ 
and $P_b=({\pp_b}_1,\ldots,{\pp_b}_N)$ are ``ghost variables'', $G_{a,b}$ is the normalization 
factor of two coherent states as defined in Eq. (\ref{eq:G}), and 
$\psi_\alpha^a(\qq)\equiv\psi^a(\qq_\alpha)$ is the coherent 
state as defined in Eq. (\ref{eq:csq}). In our notation,
\bq
\substack{\beta\\\mathlarger{\mathlarger{\mathlarger{\mathlarger{\textlinb{\Bo}}}}}\\t,t'=0}f(t,t')=\lim_{\tau\to 0}\prod_{i=0}^{M-1}f(i\tau,(i+1)\tau),
\eq
is a limit productorial over the continuous imaginary time from 0 to $\beta$. Notice that the 
antisymmetrization is only necessary on the real variables: the particles positions, as usual.
Also, is important to impose \orange{imaginary time and spatial periodic boundary conditions on the 
positions of the real and on the ghost particles, but nothing should be done on the momenta 
of the ghost particles.} 
\blue{In any case the ghost particles are always naturally bound to the real particles due 
to the effect of the HO, so their momenta never reach infinity.}
\orange{All distances between real particles and between real and ghost particles should be 
computed using the minimum image convention. This affects both the calculation of the 
potential energy between real particles and also the harmonic binding between the ghost and real 
particles enforced by the gaussian in Eq. (\ref{eq:csq}) and the phase factors in those coherent 
states. Note that while the ghost ``a'' acts at time $t$, ghost ``b'' acts at time $t+\tau$. 
This is the reason why it is important to impose imaginary time periodic boundary conditions 
also on the ghost coordinates.}}

From our preliminary \green{Metropolis \cite{Kalos-Whitlock}} simulation
\footnote{\green{
In our simulation we choose as transition move a uniform displacement of each of the $3MN$ 
real path coordinate $\qq_{\alpha}(i\tau)\to\qq_{\alpha}(i\tau)+(1/2-\eta)\mathbf{\Delta}$ for 
$\alpha=1,\ldots,N$ and $i=1,\ldots,M$, where $\eta$ is a uniform pseudo-random number in $[0,1)$ 
and $\mathbf{\Delta}$ a fixed 3-dimensional vector whose magnitude is chosen so to have
acceptance ratios close to $1/2$. And of each of the $12MN$ ghost path 
canonical variables
$\qq_{a_\alpha}(i\tau)\to\qq_{a_\alpha}(i\tau)+(1/2-\eta)\mathbf{\Delta}$, 
$\pp_{a_\alpha}(i\tau)\to\pp_{a_\alpha}(i\tau)+(1/2-\eta)\mathbf{\Delta}$, and
$\qq_{b_\alpha}(i\tau)\to\qq_{b_\alpha}(i\tau)+(1/2-\eta)\mathbf{\Delta}$, 
$\pp_{b_\alpha}(i\tau)\to\pp_{b_\alpha}(i\tau)+(1/2-\eta)\mathbf{\Delta}$. So that the 
transition probability density is just a constant and drops out of the acceptance probability.
A MC step consists of displacing the $M$ timeslices of a randomly chosen ghost and real 
particle one by one and of a displacement of a fixed random number $\leq M$ of timeslices, 
connecting two randomly chosen real particles, all at once, in a sort of ``brownian bridge'', 
so to realize a swap of two real particles. Here it is also necessary to copy the same 
bridge also for the ``a'' and ``b'' ghosts and swap them together with the real particles.}}
of distinguishable Boltzmann free 
particles in 3 dimensions we found good agreement for the kinetic energy, 
$\langle\hat{T}\rangle = 3N/2\beta$,
\footnote{\label{foot:ke-id-boltzmann}\green{
The result taking care of finite size, $L$, effects being:
\[
q(\beta) = \frac{\beta}{2m}\left(\frac{2\pi}{L}\right)^2,~~~
Z(\beta) = \theta_3^{3N}\left(0,e^{-q(\beta)}\right),~~~
\langle\hat{T}\rangle = - \frac{1}{Z(\beta)}\frac{dZ(\beta)}{d\beta},
\]
where $\theta_3$ is the elliptic theta function of third kind.
}}
calculated, \blue{for a single link of the ghost variables, from 
$\langle\hat{T}\rangle=m(\partial {\cal Z}/\partial m)/{\cal Z}\beta$, where 
${\cal Z}={\rm tr}(\hat{\rho})$ is the partition function
\footnote{\orange{
There exist at least other three estimators for the kinetic energy \citep{Ceperley1995}
that are: i. The {\sl direct} estimator obtained by taking the average of the laplacian with
the coherent states density matrix of Eq. (\ref{eq:eureka}); 
ii. the {\sl thermodynamic} estimator obtained by 
taking the partial derivative respect to the mass or to $\beta$; iii. the {\sl virial} estimator 
obtained from the virial theorem.}}, so that
\footnote{\orange{Note that this estimator is rather computationally cumbersome as it requires 
the difference of two terms that diverge in the continuum $\tau\to 0$ limit.}}
\bq \label{eq:ke}
\langle\hat{T}\rangle = \frac{3N}{2\tau}\frac{\varphi}{(1+\varphi)}\phi - \orange{\frac{\langle P_a^2(i\tau)+P_b^2(i\tau)\rangle}{4m}},
\eq
where we used the result of Eq. (\ref{eq:diag}) with
\orange{
\bq \label{eq:phi}
\varphi&\equiv &\frac{1}{2}\left(\frac{\sigma_{pw}}{\sigma_{cs}}\right)^2
=\frac{m_{h.o.}\omega\tau}{2m}
=\frac{\xi\tau}{m},\\
\sigma^2_{pw}&\equiv &2\lambda\tau,\\
\sigma^2_{cs}&\equiv &1/m_{h.o.}\omega,
\eq
}
an adimensional constant and inserted a multiplicative factor $\phi$ to the 
first term of Eq. (\ref{eq:ke}). This kinetic energy estimator reduces to the one 
of the standard plane waves path integral, in the $\xi\to\infty$ limit and for $\phi=1$. 
\orange{We see then that the characteristic parameter $\varphi$ of our CSPIMC algorithm 
is just $1/2$ times the ratio of the variance of the gaussian in the plane waves case, 
$\sigma^2_{pw}$, of Eq. (\ref{eq:Kpw}) and the one of the gaussian in the coherent 
states case, $\sigma^2_{cs}$, of Eq. (\ref{eq:csq}). We will see below that in order to have 
the correct measure of the internal potential energy it is necessary to take the $\tau\to 0$ 
continuum limit but keeping $\varphi=\pi/2$ constant.}

Defining the {\sl efficiency} of the QMC algorithm as 
\orange{$\gamma_{\cal O}=1/\sigma_{\langle\cal O\rangle}{\cal T}_{\rm MC}$}, where 
${\cal T}_{\rm MC}$ is the total {\sl computer time} and $\sigma_{\langle\cal O\rangle}$ is 
the estimator for the {\sl standard deviation} of the measurement of the mean for property 
${\cal O}$, we generally find, for the calculation through Eq. (\ref{eq:ke}), 
\orange{$\gamma^{\rm cs}_{\hat{T}}/\gamma^{\rm pw}_{\hat{T}}\approx 1/6$}, where 
$\gamma^{\rm cs}$ is the efficiency of the PIMC with Coherent States and $\gamma^{\rm pw}$ is 
the efficiency of the PIMC with Plane Waves.}

\begin{figure}[htbp]
\begin{center}
\includegraphics[width=8cm]{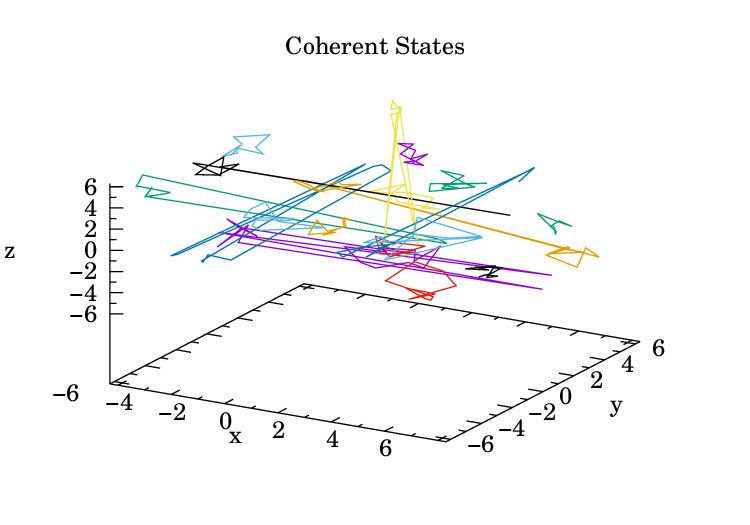}
\includegraphics[width=8cm]{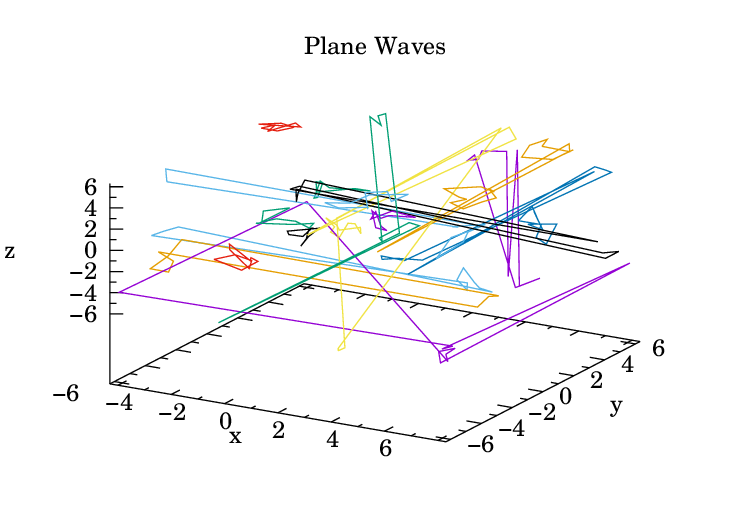}
\end{center}  
\caption{We show a comparison between equilibrium snapshots of \blue{real paths of} a simulation for 
free distinguishable particles with our coherent states kinetic energy {density matrix} (left panel) 
and one with the standard plane waves \citep{Ceperley1995} {density matrix} (right panel) for a test 
case of $N=20, m=1, N/L^3=0.01, T=0.3, m_{h.o.}\omega=0.7, M=10$ for which \blue{we found 
$\langle\hat{T}\rangle=9.0(7)$ with calculation (\ref{eq:ke}) with $\phi\approx 6.5$ and the exact 
result is} \green{$\langle\hat{T}\rangle\approx 8.99999996090802053$}. Different (real) paths have 
different colors. \orange{In our CSPIMC the ghost paths are not shown but they 
dance nearby the real paths to which they are bound by the Gaussian of Eq. (\ref{eq:csq}). 
A snapshot of the ghost paths is therefore very similar to the snapshots of the real paths. 
The ghost momenta at equilibrium are just centered spherically around the origin with an extent 
that shrinks upon choosing higher $\xi$ values. See for example Fig. \ref{fig:cspimc-lj}.}} 
\label{fig:paths}
\end{figure}

In Figure \ref{fig:paths} we show snapshots of the real paths in each of the two schemes for a test 
case of $N=20$ free distinguishable particles of unit mass $m=1$, with a density $N/L^3=0.01$, at a 
temperature $T=0.3$, and $m_{h.o.}\omega=0.7, M=10$. 
\orange{We see that the real path from our CSPIMC simulation are very similar to the ones of the 
standard plane wave PIMC. The result for the kinetic energy of Eq. (\ref{eq:ke}) is comparable and 
agrees well with the thermodynamic limit exact result of $9$, giving $9.0(7)$ after $10^6$ MC steps 
for $\phi\approx 6.5$.}

\green{Choosing higher values for the parameter $2\xi=m_{h.o.}\omega$ slows down the simulation: 
In order to have acceptance ratios close to $1/2$ it becomes necessary to choose smaller and smaller 
displacements $\Delta$.
\footnote{\blue{Actually the simulation ``slows'' down both at very high and at very low 
$\xi$.}}
But, in agreement with the discussion in Appendix \ref{app:zeta}, allows to use larger timesteps, 
which speeds up the convergence towards the continuum limit. Note that choosing higher values 
of $\xi$ produces more tightly bound real paths to the ghost particles. \orange{It is important to 
stress that the ghost coordinates are coupled harmonically through gaussians 
(see Eq. (\ref{eq:csq})) to the real coordinates and it is important to employ the minimum 
image convention when calculating the distance between the ghost particle and the real particle 
that enters the exponent of the gaussian. In Table \ref{table} we performed some test 
CSPIMC simulations for the case of $N=20$ distinguishable free particles with 
$m=1, N/L^3=0.01, T=0.3, M=10$ at fixed $\tau=1/3$ and increasing values of $\xi$ in order to 
determine the behavior of the factor $\phi$. We see that $\phi$ tends towards the limiting value 
$\phi_\infty\approx 2$ valid for that particular $\tau$. We will generally have 
$\phi_\infty=\phi_\infty(\tau)$. Moreover at a fixed value of $\varphi$ the simulation does not 
depend on the thermodynamic state, $\rho, T$, apart from the finite size and the continuum limit 
effects. \brown{We also verified that in the continuum limit, 
$\tau\to 0$, but at fixed $\varphi$, $\phi$ tends towards $\approx\sqrt 3$, in Eq. (\ref{eq:ke}), 
as it is shown in Table \ref{table1} for the particular case of $\varphi=\pi/2$}. 
\footnote{\brown{Note that this does not hold for the interacting system, as it is shown in the 
supplementary material. This is due to the fact that in the estimator of the kinetic energy 
there are terms depending on the interaction as shown in Ref. \cite{Ceperley1995}.}}
}

\begin{table}[hbt]
\caption{For a test case of $N=20$ distinguishable free particles with 
$m=1, N/L^3=0.01, T=0.3, M=10$, for which the finite size exact thermodynamic result for the 
kinetic energy is $\langle\hat{T}\rangle\approx 8.99999996090802053$, we determined the factor 
$\phi$ in Eq. (\ref{eq:ke}) for increasing values of $2\xi=m_{h.o.}\omega$. One can see that 
$\phi\to\phi_\infty(\tau=1/3)\approx 2$. In the simulations we used up to $2\times 10^6$ MC steps.
\orange{The parameter $\varphi=\xi\tau/m$, whose importance for our CSPIMC has been stressed 
in the text, is also reported in the table}.} 
\label{table}
\begin{center}
\orange{
\begin{tabular}{||c||c||c||c||}
\hline
$2\xi$ & $\varphi$ & $\langle P_a^2(i\tau)+P_b^2(i\tau)\rangle/4m$ & $\phi $ \\
\hline
\hline
7/10  & $7/60$ & 52.2(4)  & 6.51\\
1     & $1/6$  & 55.6(4)  & 5.02\\
2     & $1/3$  & 59.6(4)  & 3.05\\
3     & $1/2$  & 64.6(4)  & 2.45\\
4     & $2/3$  & 67.7(4)  & 2.13\\
5     & $5/6$  & 70.1(4)  & 1.93\\
$3\pi$& $\pi/2$& 75.5(4)  & 1.54\\
10    & $5/3$  & 77.6(4)  & 1.54\\
50    & $25/3$ & 150.5(4) & 1.98\\
100   & $50/3$ & 157.5(4) & 1.96\\
500   & $250/3$& 164.5(4) & 1.95\\
\hline
\end{tabular}
}
\end{center}
\end{table}
\begin{table}[hbt]
\caption{\brown{For a test case of $N=20$ distinguishable free particles with 
$m=1, N/L^3=0.01, T=0.3$ we determined the factor $\phi$ in Eq. (\ref{eq:ke}) 
for increasing values of $M$ at fixed $\varphi=\pi/2$ and using for 
$\langle\hat{T}\rangle$ the finite size exact result of footnote
\ref{foot:ke-id-boltzmann}. One can see that, in the continuum $\tau\to 0$ limit, 
$\phi\to\approx\sqrt 3$. The last entry reached equilibrium only after $2\times 10^8$ 
MC steps.}} 
\label{table1}
\begin{center}
\brown{
\begin{tabular}{||c||c||c||}
\hline
$M$ & $\langle P_a^2(i\tau)+P_b^2(i\tau)\rangle/4m$ & $\phi$ \\
\hline
\hline
10    & 75.5(4)   & 1.54\\
100   & 827.3(4)  & 1.52\\
200   & 1906.8(5) & 1.74\\
300   & 2867.7(5) & 1.74\\
500   & 4798.7(5) & 1.75\\
\hline
\end{tabular}
}
\end{center}
\end{table}

We performed preliminary Monte Carlo simulations for identical Bose free particles 
which are still exact
\footnote{\label{foot:ke-id-bose}\green{
In this case the expected kinetic energy in the thermodynamic limit is given by 
\cite{LandauCTP5,FeynmanFIP}
\[
\langle\hat{T}\rangle = \frac{3L^3}{2\beta\lambda_{dB}^3}
\left\{\begin{array}{ll}
{\rm Li}_{3/2 + 1}(1) & T<T_c\\
{\rm Li}_{3/2 + 1}(\bar{z}) & T>T_c
\end{array}\right.
\]
where ${\rm Li}_n(z)$ is the polylogarithm function, $\lambda_{dB}=\sqrt{2\pi\beta/m}$ is the de 
Broglie thermal wavelength, $T_c = (2\pi/m)[N/L^3{\rm Li}_{3/2}(1)]^{2/3}$ is the critical 
temperature for Bose-Einstein condensation, and the fugacity $\bar{z}$ is given by 
$N/L^3 = {\rm Li}_{3/2}(\bar{z})/\lambda_{dB}^3$. Note that for Fermi free particles one just 
needs to replace ${\rm Li}_{3/2}(z)\to -{\rm Li}_{3/2}(-z)$ and there is no condensation.}}, 
within \blue{the ultraviolet cutoff $\tau$}, the finite size effects, and the statistical errors, 
but require permutations sampling, which we implemented with a Monte Carlo move consisting of a 
swap of a pair of real particles chosen at random and of the corresponding ``a'' and ``b'' ghosts. 
\footnote{Note that the swap move will never change the {\sl flux of permutations} 
through the boundary conditions in a system with more than 2 particles. So it will
be impossible to measure quantities like the superfluid fraction \cite{Ceperley1995}.}
\blue{
For example, with calculation (\ref{eq:ke}), for $N=20$ identical Bose free 
particles of mass $m=1$ at a density of $N/L^3=0.01$ and a temperature $T=0.3$ with 
$2\xi=2, M=10$ we found \orange{$\langle\hat{T}\rangle=7.4(7)$} with $\phi\approx 3.03$ 
(compare with the values of Teble \ref{table}) after $10^6$ MC steps, when the 
exact thermodynamic result is $\langle\hat{T}\rangle\approx 7.4468550673806296256$.} 
On the other hand identical Fermi particles are subject to the sign problem \cite{Ceperley1991} and 
one needs to resort to some approximation even numerically with Monte Carlo. This can be realized  
with algorithms like for example the {\sl restricted path integral} method \cite{Ceperley1996}.
}

\begin{figure}[htbp]
\begin{center}
\includegraphics[width=8.5cm]{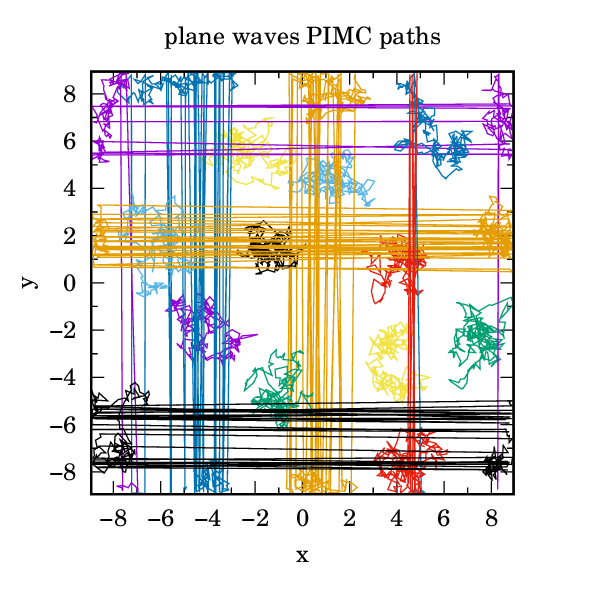}\hspace{-.5cm}
\includegraphics[width=8.5cm]{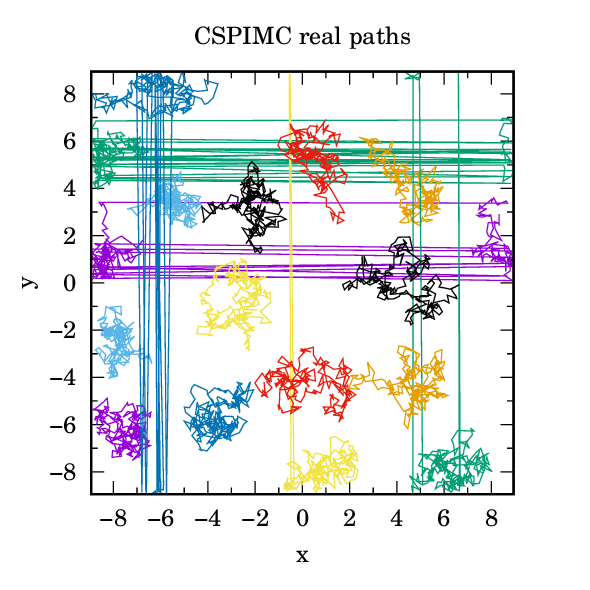}\\
\includegraphics[width=8.5cm]{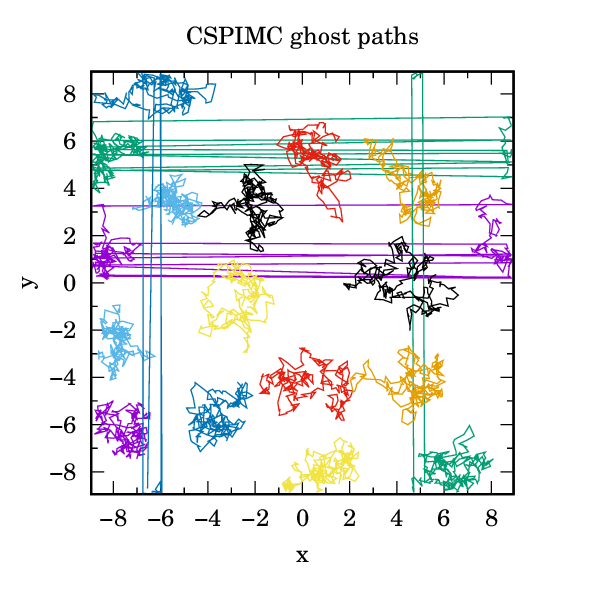}\hspace{-.5cm}
\includegraphics[width=8.5cm]{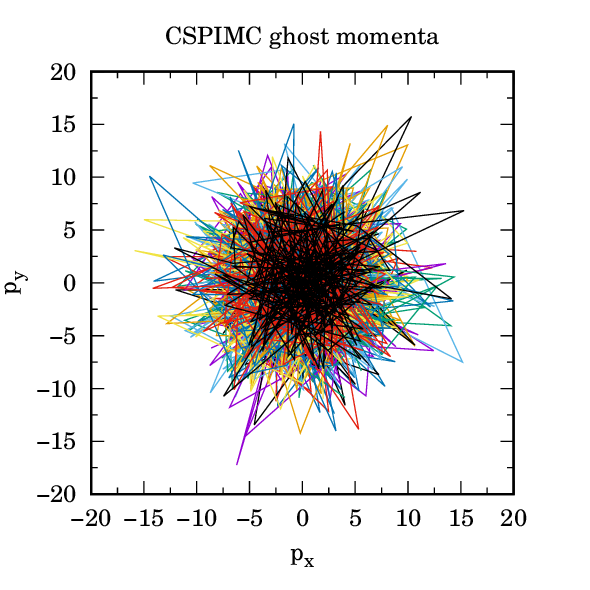}
\end{center}
\caption{\orange{We show a snapshot at equilibrium for the paths in the plane wave PIMC and the 
real paths, ghost paths, and ghost momenta for the CSPIMC for $N=16$ $^4{\rm He}$ atoms 
($m=0.08306\approx 1/12~\mbox{\AA}^{-2}\mbox{K}^{-1}$) in two dimensions interacting with a 
Lennard-Jones pair potential with parameters \cite{McMillan1964} 
$\sigma=2.556~\mbox{\AA}, \varepsilon=10.22~\mbox{K}$ and a cutoff distance 
$r_{\rm cut}=2.5~\mbox{\AA}$, at a density $N/L^2=0.05~\mbox{\AA}^{-2}$ and a temperature 
$T=1.0~\mbox{K}$ with $M=250$, $2\xi=40$, $\varphi=24/25$ in Boltzmann statistics. Different 
colors refer to different real or ghost atoms.}} 
\label{fig:cspimc-lj}
\end{figure}

\blue{We also performed some preliminary simulations with distinguishable Boltzmann 
and indistinguishable Bose particles interacting with a hard sphere, square-well \cite{Fantoni14c},
harmonic, Lennard-Jones \cite{Fantoni14c,Fantoni14d,Fantoni15b,Fantoni16a}, and Coulomb 
\cite{Fantoni13g,Fantoni18a,Fantoni18c,Fantoni21b,Fantoni21i,Fantoni23a} pair potentials (among real 
particles only). \orange{For example for $N=16$ $^4{\rm He}$ atoms 
($m=0.08306\approx 1/12~\mbox{\AA}^{-2}\mbox{K}^{-1}$)
\footnote{Here we use units where $\hbar=k_B=1$ so that the imaginary time has dimensions of 
temperature.} in two dimensions, a Lennard-Jones with parameters \cite{McMillan1964} 
$\sigma=2.556~\mbox{\AA}, \varepsilon=10.22~\mbox{K}$ and a cutoff distance 
$r_{\rm cut}=2.5~\mbox{\AA}$ (so that $v(r)=0$ for $r>r_{\rm cut}\sigma$ without 
long range corrections), at a surface density $N/L^2=0.05~\mbox{\AA}^{-2}$ and a 
temperature $T=1.0~\mbox{K}$ with $M=250$, with Boltzmann statistics and  
calculation (\ref{eq:ke}) with $2\xi=40, \varphi=24/25$, we found 
$\langle\hat{T}\rangle=83(5)~\mbox{K}$ if $\phi\approx 1.99$ (compare with the values of Teble \ref{table}) 
and $\langle\hat{V}\rangle=-83.6(3)~\mbox{K}$ throwing away the first $6\times 10^6$ equilibration 
MC steps and averaging over the last $2\times 10^6$ steps.
\footnote{\blue{
Here we chose $|\mathbf{\Delta}|=\sqrt{2}L/100$ with acceptance ratios of $\approx 0.5$.}}
This should be compared with the result from conventional plane waves PIMC \cite{Gordillo1998} 
for which we find $\langle\hat{T}\rangle=82.1(7)~\mbox{K}$ and 
$\langle\hat{V}\rangle=-83.8(1)~\mbox{K}$ where we
throw away the first $10^6$ equilibration MC steps and average again over the last $2\times 10^6$.
The statistical error on the potential energy is comparable in the two cases even if
the computer time to reach a given number of steps is bigger for the CSPIMC simulation also due 
to the necessity to use complex arithmetic.
Note that choosing larger values for $\xi$ produces a decrease of the internal 
potential energy. In order to find an even better agreement between the two calculation it 
would be necessary to increase further $\xi$ but keeping $\varphi=\pi/2$ constant in the 
CSPIMC simulation. This requires going closer to the continuum limit $\tau\to 0$ which will
necessarily be more computationally demanding.
In Fig. \ref{fig:cspimc-lj} we show a snapshot of the real and ghost variables in our CSPIMC 
simulation and compare it with the one for the standard plane waves PIMC.}}

%

\section{Conclusions}
\label{sec:conclusions}

We propose a new quantum simulation method for a many body liquid. The method creates a bridge 
between coherent states (CS) 
\cite{KlauderJCP1963a,KlauderJCP1963b,KlauderJCP1963c,KlauderJCP1963d,KlauderJCP1963e,Klauder-CS} and 
conventional Path Integral Monte 
Carlo (PIMC) \cite{Ceperley1995} merged together into a Coherent States Path Integral Monte Carlo 
(CSPIMC) method. The idea hinges upon expanding the high temperature density matrix on the 
overcomplete set of single particles coherent states of John Rider Klauder 
\cite{KlauderJCP1963a,KlauderJCP1963b,KlauderJCP1963c,KlauderJCP1963d,KlauderJCP1963e,Klauder-CS}. As 
the {\sl stiffness}, \green{$\xi\equiv m_{h.o.}\omega/2$}, of the subtending harmonic oscillator (HO)
varies from low values to very high values the coherent states probability distribution changes
from Gaussian to Dirac delta like. 
We believe that going towards a more and more stiff HO the resulting extremely
spiked coherent states could render the Quantum Monte Carlo (QMC) simulation more and more efficient 
since one is allowed to work with larger timesteps.

We are often interested in the ensemble thermal average 
$\langle\hat{{\cal O}}\rangle={\rm tr}(\hat{\rho}\hat{{\cal O}})/{\rm tr}(\hat{\rho})$ of an 
observable ${\cal O}$ at a given finite inverse temperature $\beta$. Using the coordinate 
representation for the density matrix $\hat{\rho}$, as in Eq. (\ref{eq:trotter}), we arrive to 
the sought for path integral expression \green{with closed paths in periodic imaginary time, such 
that $t+\beta=t$ which must be enforced due to the trace in the thermal average}. 
A key ingredient is the high temperature density matrix at a small inverse temperature $\tau$. 
This is made up of two pieces: a {\sl new} kinetic energy \blue{action} and the usual diagonal
potential energy \blue{action}. We find the explicit analytic primitive approximation form of the 
kinetic energy \blue{action}. Being \blue{the kinetic energy density matrix} a product of single 
particles kinetic energy operators it is possible to expand it in the overcomplete set of single 
particles Klauder coherent states. The result is summarized into the $\zeta$ function of Eqs. 
(\ref{eq:azeta}). In particular we see from that equation that is necessary to introduce some 
``ghost variables'' along the path integral whose expression becomes the one in \red{Eq. 
(\ref{eq:eureka}). 
From this equation we see how the action (compare with Eq. (1.1) of Ref. \cite{Ceperley1996}) now 
has a kinetic energy part \orange{where the ghost variables (coordinates and momenta) play a role}
and the potential energy part that remains the one of the real particles, as usual. 
Therefore it is important to impose periodic \orange{boundary conditions on imaginary time and 
on space both for the real and ghost coordinates but not on the ghost 
momenta. Of course the distance between particle-particle and ghost-particle should also be 
computed using the minimum image convention.}}

Our calculation shows that for a very stiff HO the high temperature kinetic energy density matrix 
tends to become extremely peaked \green{(see Fig. \ref{fig:ztl})}. We suggest that this could allow
an improvement in the efficiency $\gamma_{\cal O}$.
\red{Of course it is necessary to choose $\tau$ small enough so that our approximation 
(\ref{eq:csapprox}) does not break down.} 
\green{Our preliminary simulations \blue{on distinguishable free particles,
(see Fig. \ref{fig:paths}) confirm this. In particular we showed that increasing $m_{h.o.}\omega$ 
the result for the kinetic energy tends to the correct limiting value in accord with the
exact finite size result from thermodynamics.}
\blue{So we are facing a compromise between the slowing down due to the choice of 
a higher $m_{h.o.}\omega$ and the speeding up due to the opportunity to choose larger 
timesteps $\tau$. It could be possible to have comparable efficiencies by choosing 
larger timesteps. Nonetheless, our new algorithm still results less 
efficient than the standard one employing plane waves and with longer 
equilibration times. Even if the increase in the computer time that we face in our CSPIMC respect 
to the standard plane waves PIMC is due to the necessity to use complex arithmetic in the former.}

So, summarizing, in our new CSPIMC scheme we require as usual the two limiting procedures:
i. thermodynamic limit, $L\to\infty$ with the density $N/L^3$ kept constant; ii. continuum limit,
$M\to\infty$ with the temperature $T=1/k_BM\tau$ kept constant
\footnote{\blue{Of course this limit is superfluous for a non interacting system where one 
always just needs two timeslices.}} 
\blue{But the novelty respect to more conventional path integral methods is the new degree of 
freedom offered by the mass, $m_{h.o.}$, and elastic constant, $k=m_{h.o.}\omega^2$, 
of the harmonic oscillator defining the choerent states overcomplete basis set. In the limit of 
a very stiff harmonic oscillator, $\xi\equiv m_{h.o.}\omega/2\to\infty$, the harmonic oscillator ghost 
coordinates coincides with the real particles coordinates and CSPIMC reduces to conventional 
plane waves PIMC. For finite values of $m_{h.o.}\omega$ the real particles are elastically bound 
to the ghost particles and they are mutually influenced: The ghost pulls the particle and the 
particle pushes the ghost! In the action the kinetic energy is that of the ghosts but the 
potential energy is that among the particles. \orange{In approaching the continuum limit it is 
{\sl necessary} to keep $\varphi\equiv\xi\tau/m = \pi/2$ constant in order to have the correct 
internal potential energy measure, $\varphi$ being the characteristic parameter for our 
CSPIMC algorithm. We then see that with our algorithm we can relax the continuum $\tau\to 0$ limit
by choosing $\xi\to\infty$.
}}}

\blue{We will worry about the mathematical rigor of the primitive approximation proposed here, 
as described in Appendix \ref{app:zeta}, or 
about its refinements \cite{Ceperley1995}, in future \red{more specialized} works. 
This could be important also to further increase the computational efficiency of the 
numerical experiment.}

\brown{
The use of Coherent States in a quantum many body calculation could eventually lead to
the formulation of a quantum version of a molecular dynamics simulation. Klauder in Ref. 
\cite{KlauderJCP1963b} determines generalized relation between quantum and classical 
dynamics using the coherent states formalism that he calls continuous representation.
As is shown in Section 2.B of that reference such correspondence is only inexact if the 
Hamiltonian is not linear or quadratic in the canonical variable. But some progress can be 
made for Hamiltonians where the potential energy is just a function of the positions. This
was the main motivation that triggered the development of the present work on a Monte Carlo
simulation instead, but we intend to reconsider the challenge of the development of a
quantum molecular dynamic computer experiment soon enough. This would result in a unifying
formalism for a quantum many body computational experiment.  

Additional support to the findings reported in this work is contained in the supplementary
material where the listing of the FORTRAN code used in the CSPIMC simulation is given.}

\begin{acknowledgments}
I would like to thank prof. John Rider Klauder for transmitting me the passion for coherent states and 
all the `family' of the Physics Department in Gainesville at the University of Florida which invited me 
to participate at the Memorial Conference held on 15 February 2025 in honor of the professor path in life 
and science. The professor passed away on 24 October 2024 at 92 years of age. I would like also to thank
prof. Saverio Moroni for his support in creating the 
computer code for the plane wave simulations for $^4$He in Bose statistics.
\end{acknowledgments}

\appendix
\section{Determination of $K$ in Eq. (\ref{eq:K})}
\label{app:K}
From Eq. (\ref{eq:pa}) and inserting the resolution of the identity from Eq. 
(\ref{eq:spe}) in terms of the complete set of coherent states two times we find
\bq
\rho(Q,Q';\tau)\approx \frac{1}{N!^2}\sum_{\aaa,\bbb}\langle Q|{\rm det}|| \,|\psi_\alpha^\aaa\rangle\langle\psi_{\beta}^\aaa|\, ||e^{-\tau\hat{T}}{\rm det}|| \,|\psi_\alpha^\bbb\rangle\langle\psi_{\beta}^\bbb|\, ||\,|Q'\rangle e^{-\tau\blue{\upsilon}V(Q')},
\eq
here
\bq
\sum_{\aaa,\bbb}\ldots \to\int \ldots \frac{d\qq_a\,d\pp_a}{(2\pi)^3}\frac{d\qq_b\,d\pp_b}{(2\pi)^3}.
\eq
Now the two antisymmetrizations are redundant and one can safely keep just one of the two. 
We then find
\bq \nonumber
\rho(Q,Q';\tau)&\approx&\frac{1}{N!}\sum_{\cal P}(-1)^{\cal P}\sum_{\aaa,\bbb}\prod_{\alpha,\beta}\langle\qq_{1},\ldots,\qq_{N}|\psi_{{\cal P}\alpha}^\aaa\rangle\langle\psi_{\beta}^\aaa|e^{-\tau\hat{T}}|\psi_\alpha^\bbb\rangle\langle\psi_{\beta}^\bbb|Q'\rangle e^{-\tau\blue{\upsilon}V(Q')}\\ \label{eq:rhocs}
&=&\frac{1}{N!}\sum_{\cal P}(-1)^{\cal P}\sum_{\aaa,\bbb}\prod_{\alpha}\langle\qq_{{\cal P}1},\ldots,\qq_{{\cal P}N}|\psi_\alpha^\aaa\rangle\langle\psi_{\alpha}^\aaa|e^{-\tau\hat{T}_\alpha}|\psi_\alpha^\bbb\rangle\langle\psi_{\alpha}^\bbb|Q'\rangle e^{-\tau\blue{\upsilon}V(Q')},
\eq
where we decided to keep the antisymmetrization only on the left particles positions and in the 
last equality we used \blue{the resolution of the identity (intimately connected to the 
Segal-Bargmann transform)
$\sum_{\aaa}|\psi_\alpha^\aaa\rangle\langle\psi_{\beta}^\aaa|=\delta_{\alpha,\beta}$ }. We will
also use the following orthogonality condition among single particle coherent states
\bq
\langle\psi_{\beta}^\aaa|\psi_\alpha^\bbb\rangle=G_{\aaa,\bbb}\delta_{\alpha,\beta},
\eq
where $\delta$ is a Kronecker delta symbol and
\bq \label{eq:G}
G_{\aaa,\bbb}&=&e^{-\frac{1}{2}(|\aaa|^2+|\bbb|^2)+\aaa^*\cdot \bbb+\frac{i}{2}(\qq_a\cdot\pp_a-\qq_b\cdot\pp_b)},\\ \label{eq:i}
\aaa&=&\frac{1}{\sqrt{2m_{h.o.}\omega}}(m_{h.o.}\omega\qq_a+i\pp_a),\\
\bbb&=&\frac{1}{\sqrt{2m_{h.o.}\omega}}(m_{h.o.}\omega\qq_b+i\pp_b),
\eq
here $\omega=\sqrt{k/m_{h.o.}}$ is the angular frequency of the harmonic oscillator of elastic 
constant $k$ and mass $m_{h.o.}$.

We then find from Eq. (\ref{eq:K})
\bq \label{eq:KK}
e^{-K(Q,Q';\tau,m,\xi)}&=&\frac{1}{N!}\sum_{\cal P}(-1)^{\cal P}\prod_{\alpha}\sum_{\aaa,\bbb}\langle\qq_{{\cal P}1},\ldots,\qq_{{\cal P}N}|\psi_\alpha^\aaa\rangle\langle\psi_\alpha^\aaa|e^{-\tau\hat{\pp}^2_\alpha/2m}|\psi_{\alpha}^\bbb\rangle\langle\psi_{\alpha}^\bbb|\qq_{1}',\ldots,\qq_{N}'\rangle\\ \label{eq:zeta}
&\equiv&\frac{1}{N!}\sum_{\cal P}(-1)^{\cal P}\prod_{\alpha}\zeta_\alpha\Bigl[\qq_{{\cal P}\alpha}|\qq'_\alpha;\tau,m,\xi\Bigr],
\eq
where $\langle Q|\psi^\aaa\rangle=\langle Q|\qq_a,\pp_a\rangle$ is the 
position representation of the single particle coherent state. The element 
$\langle\psi_{\alpha}^\aaa|e^{-\tau\hat{T}_\alpha}|\psi_\alpha^\bbb\rangle$ is found in 
Appendix \ref{app:ome}. The function $\zeta$ is defined 
in Eq. (\ref{eq:zeta}) and determined in Appendix \ref{app:zeta}.

\blue{It is important to extract the diagonal part of the high temperature kinetic density matrix
of the distinguishable particles. From Eq. (\ref{eq:KK}) and using the approximation of 
Eq. (\ref{eq:prop}) and the definitions of Eqs. (\ref{eq:csq}) and (\ref{eq:G}) we readily find
\bq \label{eq:diag}
e^{-K(Q,Q;\tau,m,\xi)}=\left[\frac{\xi}{2\sqrt{\pi\xi(1+2\xi\lambda\tau)}}\right]^{3N},
\eq
where $\xi=m_{h.o.}\omega/2$. Note that in the $\xi\to\infty$ limit we find the usual result for 
plane waves $(4\pi\lambda\tau)^{-3N/2}$ but times $2^{-3N/2}$.
}

\section{Calculation of the element of Eq. (\ref{eq:KK})}
\label{app:ome}

We want here calculate explicitly the matrix element of Eq. (\ref{eq:KK}).
We can then think that we find
\bq \label{eq:csapprox}
\langle\psi_{\alpha}^\aaa|e^{-\tau\hat{T}_\alpha}|\psi_\alpha^\bbb\rangle&\approx&
\langle\psi_{\alpha}^\aaa|\psi_\alpha^\bbb\rangle e^{-\tau\orange{(\pp_a^2+\pp_b^2)/4m}}\\ \label{eq:prop}
&=&G_{\aaa,\bbb}e^{-\tau\orange{(\pp_a^2+\pp_b^2)/4m}},
\eq
where in the second equality we used the exact normalization factor of Eq. (\ref{eq:G}).
\orange{Note that the symmetrization in the kinetic energy at the exponent is crucial in the
Monte Carlo. Also we decided to use the ``a'' and ``b'' momenta at the same imaginary time
unlike the two ghosts coordinates where ``a'' is at time $t$ and ``b'' at time $t+\tau$.}
The approximation in Eq. (\ref{eq:csapprox}) is rather suggestive and extremely \red{simple}. 
We \red{believe} that it \red{will be} ``washed away'' for small $\tau$ \red{as our 
preliminary simulation confirms}. In any case this is a rather 
subtle issue that still asks for mathematical rigor. \red{Mathematically} we need to 
\red{prove} that this high temperature matrix element should be able to reconstruct the 
finite temperature density matrix as an exact path integral \cite{Schulman}.

\section{Determination of $\zeta$}
\label{app:zeta}

\begin{figure}[htbp]
\begin{center}
\includegraphics[width=8cm]{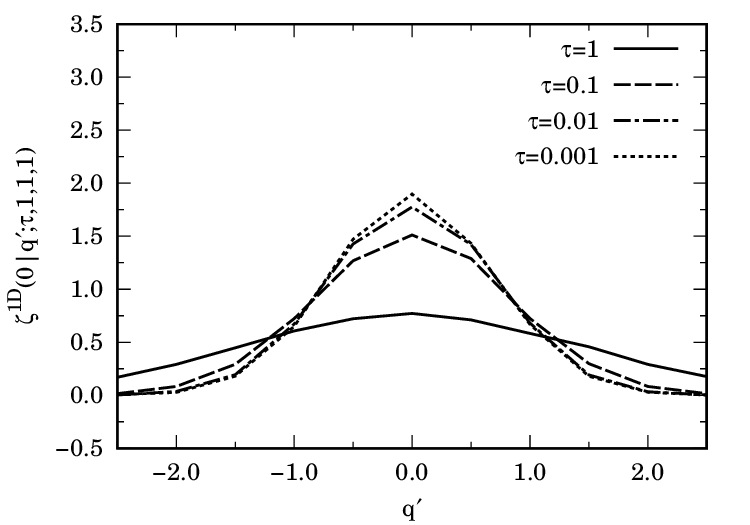}
\includegraphics[width=8cm]{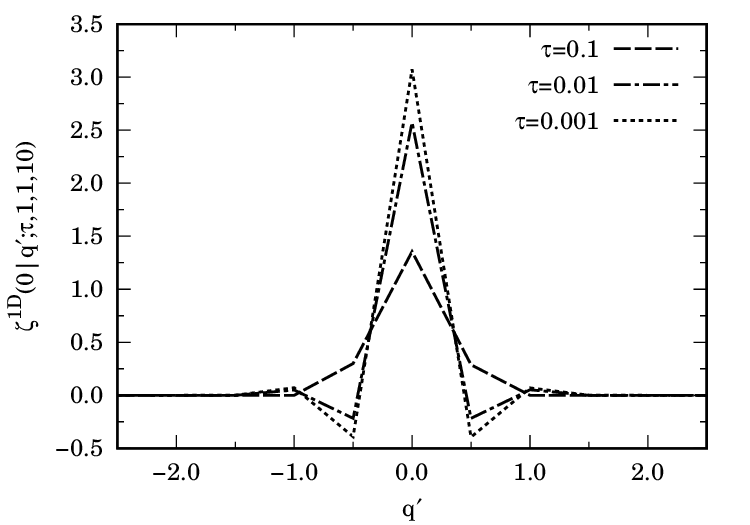}
\end{center}  
\caption{We show the one dimensional $\zeta^{1D}(q|q';\tau,m,\xi)$ calculated with 
a Monte Carlo scheme using $10^7$ (with the related statistical error) sampling points and 
choosing smaller and smaller timestep $\tau$. All the four integrations over the ghost 
variables $(q_a,p_a)$ and $(q_b,p_b)$ were chosen in the interval $[-l,l]$, $l=10$
(with the related finite size error). On the left $m_{h.o.}=1$ on the right $m_{h.o.}=10$. 
Note that the slight negative values for the high $m_{h.o.}$ and small $\tau$ case are 
numerical artifacts due to the finite size, $l<\infty$, error.} 
\label{fig:ztl}
\end{figure}

We have from the definition in Eq. (\ref{eq:zeta})
\bq \nonumber
&&\zeta_\alpha\Bigl[\qq|\qq';\tau,m,\xi\Bigr]\\
&&\equiv\sum_{\aaa,\bbb}\langle\qq|\psi_\alpha^\aaa\rangle\langle\psi_\alpha^\aaa|e^{-\tau\hat{T}_\alpha}|\psi_{\alpha}^\bbb\rangle\langle\psi_{\alpha}^\bbb|\qq'\rangle\\
&&=\int\frac{d\qq_a\,d\pp_a}{(2\pi)^3}\frac{d\qq_b\,d\pp_b}{(2\pi)^3}\psi_\alpha^\aaa(\qq)
{\psi_\alpha^\bbb}^*(\qq')\langle\psi_\alpha^\aaa|e^{-\tau\hat{T}_\alpha}|\psi_{\alpha}^\bbb\rangle\\
\label{eq:azeta}
&&\approx\int\frac{d\qq_a\,d\pp_a}{(2\pi)^3}\frac{d\qq_b\,d\pp_b}{(2\pi)^3}\psi_\alpha^\aaa(\qq)
{\psi_\alpha^\bbb}^*(\qq')G_{\aaa,\bbb}e^{-\tau\orange{(\pp_a^2+\pp_b^2)/4m}},
\eq
where $\psi_\alpha^\aaa(\qq)$ is the coordinate representation of the single $\alpha$ particle 
coherent state of Eq. (\ref{eq:csq}) and the {density matrix} 
$\langle\psi_\alpha^\aaa|e^{-\tau\hat{T}_\alpha}|\psi_{\alpha}^\bbb\rangle$ has been approximated in 
Eq. (\ref{eq:prop}) of Appendix \ref{app:ome}. Eq. (\ref{eq:azeta}) can be integrated with a 
simple Monte Carlo scheme (see Fig. \ref{fig:ztl}) and this will give us a primitive approximation for 
$\zeta$. \red{These additional integrations over $(\qq_a,\pp_a)$ and $(\qq_b,\pp_b)$ must be carried 
out at each timestep but Monte Carlo will not suffer critically since it is designed to treat highly 
multidimensional integrals. We will 
then reach a path integral both on the particles positions and on {\sl ghost variables} at two next 
time steps as shown in Eq. (\ref{eq:eureka}). In particular instead of the usual $3MN$ discretized 
multidimensional path integral we will now deal with a $5\times(3MN)$ multidimensional path integral. 
The ghost variables are the canonical pair of labels from the continuous representation of 
Klauder and can be treated as canonical variables of {\sl ghost particles} with momenta that should 
not be limited employing periodic boundary conditions.} In the $\tau\to 0$ 
limit, also within the approximation (\ref{eq:csapprox}) we will have,
\bq
\zeta_\alpha\Bigl[\qq|\qq';\tau,m,\xi\Bigr]\stackrel{\tau\to 0}{\looongrightarrow}\langle\qq|\qq'\rangle=\delta(\qq-\qq'),
\eq
where $\delta$ is a Dirac delta function. The $\tau\to 0$ limit washes away the 
$k,m_{h.o.}$ dependence. On the other hand choosing a stiff harmonic oscillator, i.e. 
one with a high mass, the $\zeta$ function width diminishes, as is shown in Fig. \ref{fig:ztl}. This 
will allow to choose bigger timesteps in the path integral, thereby reducing the computational cost.

\section*{Author declarations}

\subsection*{Conflicts of interest}
None declared.

\subsection*{Data availability}
The data that support the findings of this study are available from the 
corresponding author upon reasonable request.

\subsection*{Funding}
None declared.

\bibliography{qmd}

\end{document}